\documentclass{PoS}

\usepackage{amssymb,pslatex}
\usepackage{amsmath,marvosym}
\usepackage{braket}

\usepackage{tikz}
\usetikzlibrary{plotmarks}
\newcommand\marksymbol[2]{\tikz[#2,scale=1.2]\pgfuseplotmark{#1};}
\renewcommand{\d}{\ensuremath{\mathrm{d}}}
\newcommand{\p}{\partial}
\newcommand{\GZ}{\ensuremath{\mathrm{GZ}}}
\newcommand{\gf}{\ensuremath{\mathrm{gf}}}
\newcommand{\YM}{\ensuremath{\mathrm{YM}}}

\title{Further Study of BRST-Symmetry Breaking on the Lattice}

\ShortTitle{Lattice BRST-Symmetry Breaking}

\author{\speaker{Attilio Cucchieri}\thanks{We acknowledge
partial support from CNPq. We would like to acknowledge
computing time provided on the
Blue Gene/P supercomputer supported by the Research Computing
Support Group (Rice University) and Laborat\'orio de Computa\c c\~ao
Cient\'\i fica Avan\c cada (Universidade de S\~ao Paulo).}\\
        Instituto de F\'\i sica de S\~ao Carlos, Universidade de S\~ao Paulo,
        Caixa Postal 369, 13560-970 S\~ao Carlos, SP, Brazil\\
        E-mail: \email{attilio@ifsc.usp.br}}

\author{Tereza Mendes\\
        Instituto de F\'\i sica de S\~ao Carlos, Universidade de S\~ao Paulo,
        Caixa Postal 369, 13560-970 S\~ao Carlos, SP, Brazil\\
        E-mail: \email{mendes@ifsc.usp.br}}

\abstract{We evaluate the so-called Bose-ghost propagator $Q(p^2)$
for SU(2) gauge theory in minimal Landau gauge, considering lattice
volumes up to $120^4$ and physical lattice extents up to $13.5 \, fm$.
In particular, we investigate discretization effects, as well as the infinite-volume
and continuum limits. We recall that a nonzero value for this quantity provides direct
evidence of BRST-symmetry breaking, related to the restriction of the
functional measure to the first Gribov region.
Our results show that the prediction (from cluster decomposition) for
$Q(p^2)$ in terms of gluon and ghost propagators is
better satisfied as the continuum limit is approached.}

\FullConference{34th annual International Symposium on Lattice Field Theory\\
		24-30 July 2016\\
		University of Southampton, UK}

\begin{document}

\section{BRST-Symmetry Breaking}

The minimal Landau gauge in Yang-Mills theories
\cite{Gribov:1977wm} is obtained by restricting the functional
integral to the set of transverse gauge configurations for which the
Faddeev-Popov (FP) matrix ${\cal M}$ is non-negative, the
so-called first Gribov region $\Omega$.
On the lattice, this gauge condition is imposed by considering
a minimization procedure. On the contrary, in the Gribov-Zwanziger (GZ)
approach in the continuum \cite{Zwanziger:1991ac}, this restriction
is forced by adding a nonlocal horizon-function term
$\gamma^4 S_{\mathrm{h}}$ to the usual (Landau-gauge) action.
The resulting (nonlocal) GZ action may be
localized by introducing the auxiliary fields
$\phi^{ab}_{\mu}(x)$ and
$\omega^{cd}_{\nu}(y)$, yielding
$S_{\GZ} \, = \, S_{\YM} + S_{\gf} +
S_{\mathrm{aux}} + S_{\gamma}$. Here, $S_{\YM}$
is the usual four-dimensional Yang-Mills action,
$S_{\gf}$ is the covariant-gauge-fixing term,
$S_{\mathrm{aux}}$ is defined as
\begin{equation}
S_{\mathrm{aux}} \, = \, \int \d^{\rm 4} x \,
\Bigl[ \overline{\phi}_{\mu}^{ac} \, \p_{\nu} \left( D_{\nu}^{ab}
                     \phi^{bc}_{\mu} \right) -
     \overline{\omega}_{\mu}^{ac} \, \p_{\nu} \left( D_{\nu}^{ab}
                     \omega^{bc}_{\mu} \right)
            \, - \, g_0 \left( \p_{\nu}
         \overline{\omega}_{\mu}^{ac} \right) f^{abd} \,
                D_\nu^{be} \eta^e \phi_{\mu}^{dc} \Bigr]
       \nonumber
\end{equation}
and is necessary to localize the horizon function, and $S_{\gamma}$,
given by
\begin{equation}
 S_{\gamma} \, = \, \int \d^{\rm 4}x \,
           \Bigl[ \gamma^{2}
                 D^{ab}_{\mu} \Bigl( \phi_{\mu}^{ab}
           + \overline{\phi}_{\mu}^{ab} \Bigr)
        - 4 \left( N_c^{2} - 1 \right) \gamma^{4} \Bigr]
     \; , \nonumber
\label{eq:Sofgamma}
\end{equation}
allows one to fix the $\gamma$ parameter
through the so-called horizon condition.
Also, one can define \cite{Vandersickel:2012tz}
for these fields a nilpotent BRST transformation $s$
that is a simple extension of
the usual (perturbative) BRST transformation leaving
$S_{\YM} + S_{\gf}$ invariant.
However, in the GZ case, the BRST symmetry $s$ is
broken by terms proportional to a power of the
Gribov parameter $\gamma$.
Since a nonzero value of $\gamma$ is related
to the restriction of the functional integration to
$\Omega$,
it is somewhat natural to expect a breaking of the (extended)
BRST symmetry $s$, as a direct consequence of the
nonperturbative gauge-fixing.\footnote{This issue has been
investigated in several works (see e.g.\ \cite{Zwanziger:2009je,
brst,Capri:2015ixa,Pereira:2016inn} and
references therein).}
More precisely ---as nicely explained in Ref.\
\cite{Pereira:2016inn}---
an infinitesimal gauge transformation is formally
equivalent to a (perturbative) BRST transformation. Since
the region $\Omega$ is free of infinitesimal gauge copies,
applying $s$ to a configuration
in $\Omega$ should result in a configuration
outside $\Omega$. The breaking of the BRST symmetry
in minimal Landau gauge is then
inevitable, since the functional integration is limited to the
region $\Omega$.
This interpretation is supported by the introduction
\cite{Capri:2015ixa} of a nilpotent
nonperturbative BRST transformation
$s_{\gamma}$ that
leaves the local GZ action invariant.
The new symmetry is a simple modification of the extended BRST
transformation $s$, by adding (for some of
the fields) a nonlocal term proportional to a power of the
Gribov parameter $\gamma$.

\section{The Bose-Ghost Propagator}

As implied above,
the Gribov parameter $\gamma$ is not
introduced explicitly on the lattice, since in this case
the restriction of gauge-configuration space to the region
$\Omega$ is achieved by numerical
minimization.
Nevertheless, the breaking of the BRST symmetry $s$
induced by the GZ action may be investigated by the
lattice computation of suitable observables, such as the
so-called Bose-ghost propagator
\begin{equation}
Q^{abcd}_{\mu \nu}(x,y) \, = \,
\braket{ \, s (\, \phi^{ab}_{\mu}(x) \,
\overline{\omega}^{cd}_{\nu}(y)) \, } \, = \,
\braket{ \,
 \omega^{ab}_{\mu}(x) \, \overline{\omega}^{cd}_{\nu}(y) \, + \,
 \phi^{ab}_{\mu}(x) \, \overline{\phi}^{cd}_{\nu}(y) \, }
     \, .  \nonumber
\end{equation}
Since this quantity is BRST-exact with respect to the
(extended) BRST transformation $s$, it should be zero for
a BRST-invariant theory, but it does not necessarily vanish if the
symmetry $s$ is broken.
On the lattice, however, one does not have direct access to the auxiliary fields
$(\overline{\phi}^{ac}_{\mu}, \phi^{ac}_{\mu} )$ and
$(\overline{\omega}^{ac}_{\mu}, \omega^{ac}_{\mu} )$.
Nevertheless, these fields enter the continuum action at most
quadratically and they can be integrated out exactly, giving for the
Bose-ghost propagator an expression that is suitable for
lattice simulations. This yields 
\begin{equation}
Q^{abcd}_{\mu \nu}(x-y) \, = \, \gamma^4 \, \left\langle \,
       R^{a b}_{\mu}(x) \, R^{c d}_{\nu}(y) \, \right\rangle
    \; , \label{eq:Qprop}
\end{equation}
where
\vspace{1.5mm}
\begin{equation}
R^{a c}_{\mu}(x) = \int \d^{\rm 4} z \,
         ( {\cal M}^{-1} )^{ae}(x,z) \, B^{ec}_{\mu}(z)
\label{eq:R}
\end{equation}
and $B^{ec}_{\mu}(z)$ is given by the
covariant derivative $D^{ec}_{\mu}(z)$.
One can also note that, at the classical level, the
total derivatives
$\p_{\mu} ( \phi_{\mu}^{aa} + \overline{ \phi}_{\mu}^{aa})$
in the action $S_{\gamma}$ can be
neglected \cite{Vandersickel:2012tz,Zwanziger:2009je}. 
In this case the expression for $B^{ec}_{\mu}(z)$
simplifies to
\begin{equation}
B^{ec}_{\mu}(z) \, = \, g_0 \, f^{e b c} \, A^{b}_{\mu}(z)
  \; ,
\label{eq:Bshort}
\end{equation}
as in Ref.\ \cite{Zwanziger:2009je}. Let us stress that, in both cases,
the expression for $Q^{abcd}_{\mu \nu}(x-y)$
in Eq.\ (\ref{eq:Qprop}) depends only on the gauge field
$A^{b}_{\mu}(z)$ and can be evaluated
on the lattice.

\section{Numerical Simulations and Results}

The first numerical evaluation of the Bose-ghost propagator
in minimal Landau gauge was presented ---for the SU(2)
case in four space-time dimensions--- in Refs.\
\cite{Cucchieri:2014via,Cucchieri:2014xfa}.
In particular, we evaluated the scalar function $Q(k^2)$
defined [for the SU($N_c$) gauge group] through
the relation
\begin{equation}
Q^{a c}(k) \, \equiv \, Q^{abcb}_{\mu \mu}(k) \, \equiv \,
\delta^{a c} N_c \, P_{\mu \mu}(k) \, Q(k^2)
\nonumber \; ,
\end{equation}
where $P_{\mu \nu}(k)$ is the usual transverse projector
and $k$ is the wave vector with components $k_{\mu} = 0,
1, \ldots, N-1$, for a lattice of $N$ points per
directions. The lattice momentum $p^2(k)$ is obtained using
the improved definition (see Ref.\ \cite{Cucchieri:2014via}).
This calculation has been recently extended in Ref.\ \cite{Cucchieri:2016czg},
where we have investigated the approach to the
infinite-volume and continuum limits by considering
four different values of the lattice coupling
$\beta$ and different lattice volumes $V = N^4$,
with physical volumes ranging
from about $(3.366 \, fm)^4$
to $(13.462 \, fm)^4$.
We find no significant finite-volume effects in the data.
As for discretization effects, we observe small
such effects for the coarser lattices, especially in the
IR region.
We also tested three different discretizations\footnote{See Eqs.\
(30), (31) and (32) in Ref.\ \cite{Cucchieri:2016czg}.}
for the sources $B^{bc}_{\mu}(x)$, used in the inversion
of the FP matrix ${\cal M}$, and find that
the data are fairly
independent of the chosen lattice discretization of these sources.

\begin{figure}
\begin{center}
\includegraphics[trim=55 0 40 0, clip, scale=1.00, width=0.8\linewidth]{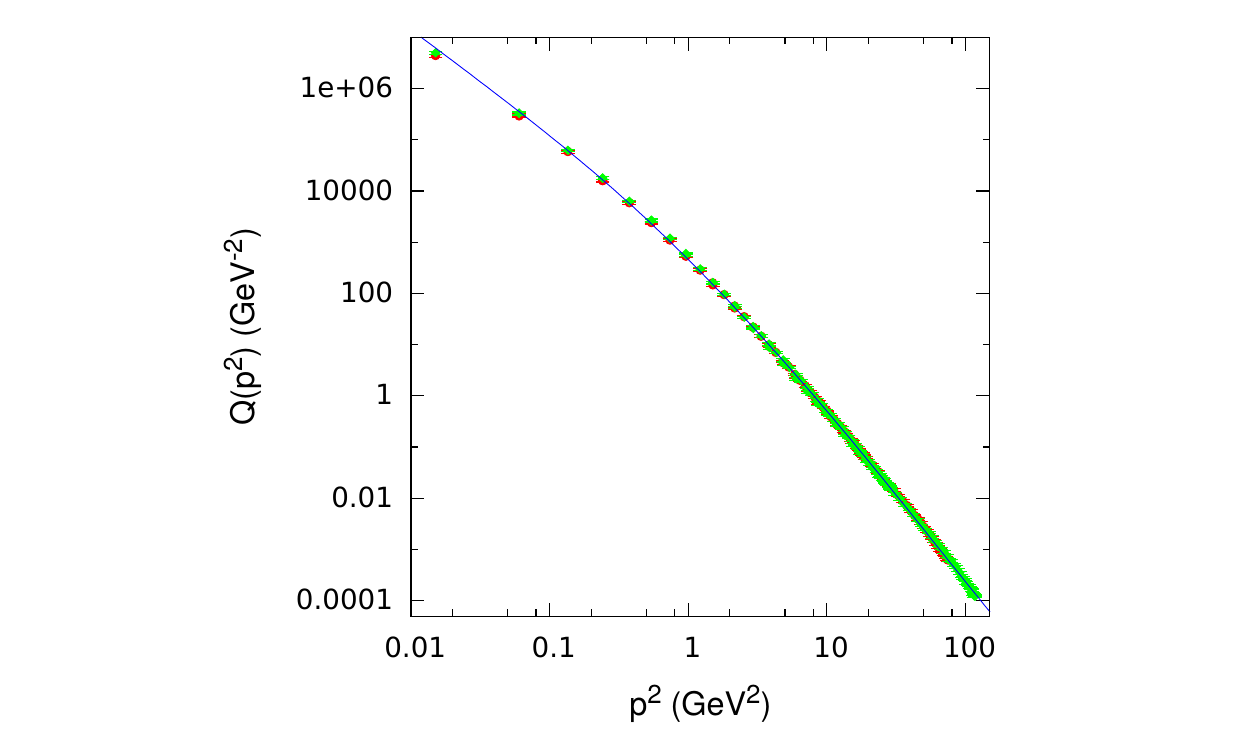}
\caption{
\label{fig:fit}
The Bose-ghost propagator
$Q(p^2)$ as a function of the (improved) lattice momentum
squared $p^2$.
Here we used as sources $B^{ec}_{\mu}(z)$
the formula reported in Eq.\ (32) of Ref.\ \cite{Cucchieri:2016czg}.
We plot data for $\beta_2 \approx 2.44$,
$V = 96^4$ (\protect\marksymbol{*}{red}) and
$\beta_3 \approx 2.51$,
$V = 120^4$ (\protect\marksymbol{diamond*}{green}),
after applying a matching procedure \protect\cite{rescaling}
to the former set of data.
We also plot, for $V = 120^4$, a fit using the
fitting function (\protect\ref{eq:fit}).
Note the logarithmic scale on both axes.
}
\end{center}
\end{figure}

Our results concerning the BRST symmetry-breaking
and the form of the Bose-ghost propagator are similar to the
previous analysis \cite{Cucchieri:2014via,Cucchieri:2014xfa}, i.e.\ we find a
$1/p^6$ behavior at large momenta and a
double-pole singularity at small momenta,\footnote{As proven in
Ref.\ \cite{Cucchieri:2016czg}, the Fourier transform of the
quantity $R^{a c}_{\mu}(x)$, defined in Eq.\ (\ref{eq:R}) above,
is trivially equal to 0 at zero momentum, i.e.\
$\sum_x \, R^{a c}_{\mu}(x) = 0$ .
Thus, one needs to consider sufficiently
large lattice volumes, in order to have the IR behavior of
the Bose-ghost propagator under control.}
in agreement with the one-loop analysis
carried out in Ref.\ \cite{Gracey:2009mj}.  
These behaviors can be clearly seen in Fig.\ \ref{fig:fit},
where we fit the data for the Bose-ghost propagator
$Q(p^2)$ using the fitting function
\begin{equation}
f(p^2) \, = \, \frac{c}{p^4} \, \frac{p^2 + s}{p^4 \, + \,
                           u^2 p^2 \, + \, t^2 }
  \; ,
\label{eq:fit}
\end{equation}
which can be related [see Eq.\ (\ref{eq:QpropDG}) below]
to an IR-free FP ghost propagator $G(p^2) \sim 1/p^2$
in combination with a massive gluon propagator $D(p^2)$.

\begin{figure}
\begin{center}
\includegraphics[trim=55 0 40 0, clip, scale=1.00, width=0.8\linewidth]{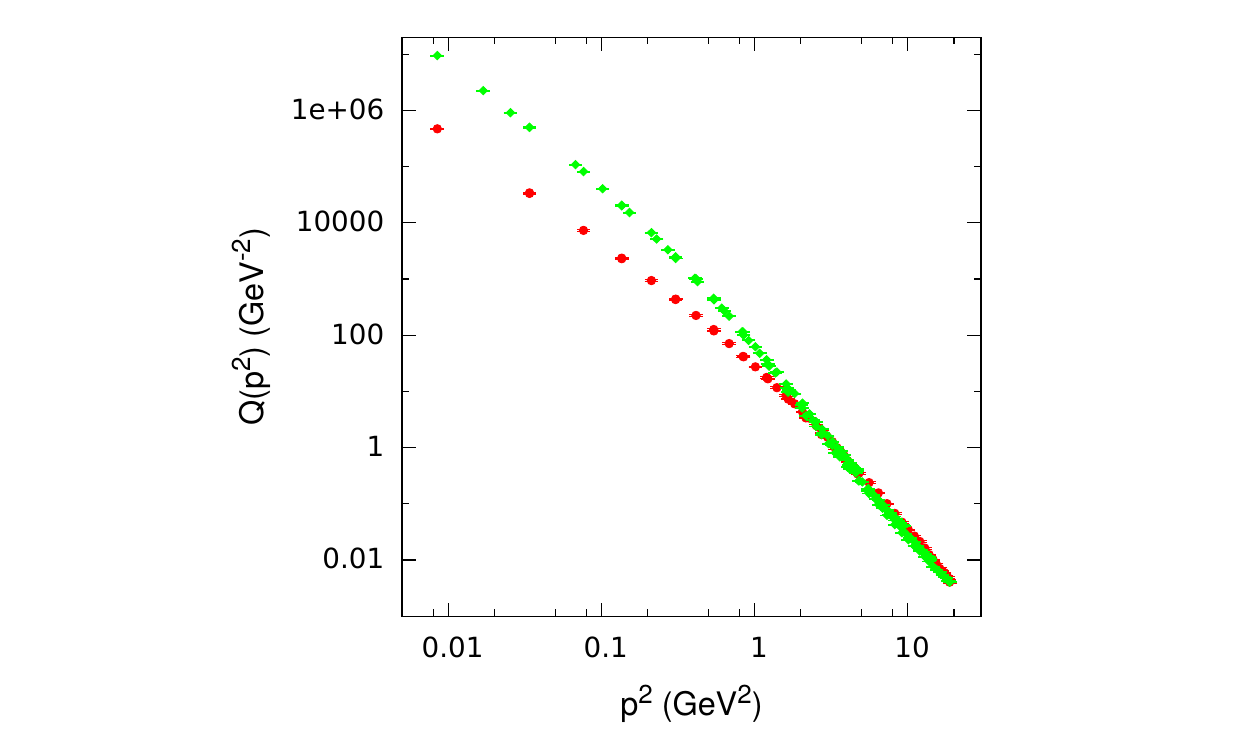}
\caption{
\label{fig:props-beta0}
The Bose-ghost propagator
$Q(p^2)$ (\protect\marksymbol{*}{red}) and the product
$g_0^2 \, G^2(p^2) \, D(p^2)$ (\protect\marksymbol{diamond*}{green})
as a function of the (improved) lattice momentum
squared $p^2$
for the lattice volume $V = 64^4$ at $\beta_0$.
Here we used as sources $B^{ec}_{\mu}(z)$
the formula reported in Eq.\ (32) of Ref.\ \cite{Cucchieri:2016czg}.
Note the logarithmic scale on both axes.
}
\end{center}
\end{figure}

\begin{figure}
\begin{center}
\includegraphics[trim=55 0 40 0, clip, scale=1.00, width=0.8\linewidth]{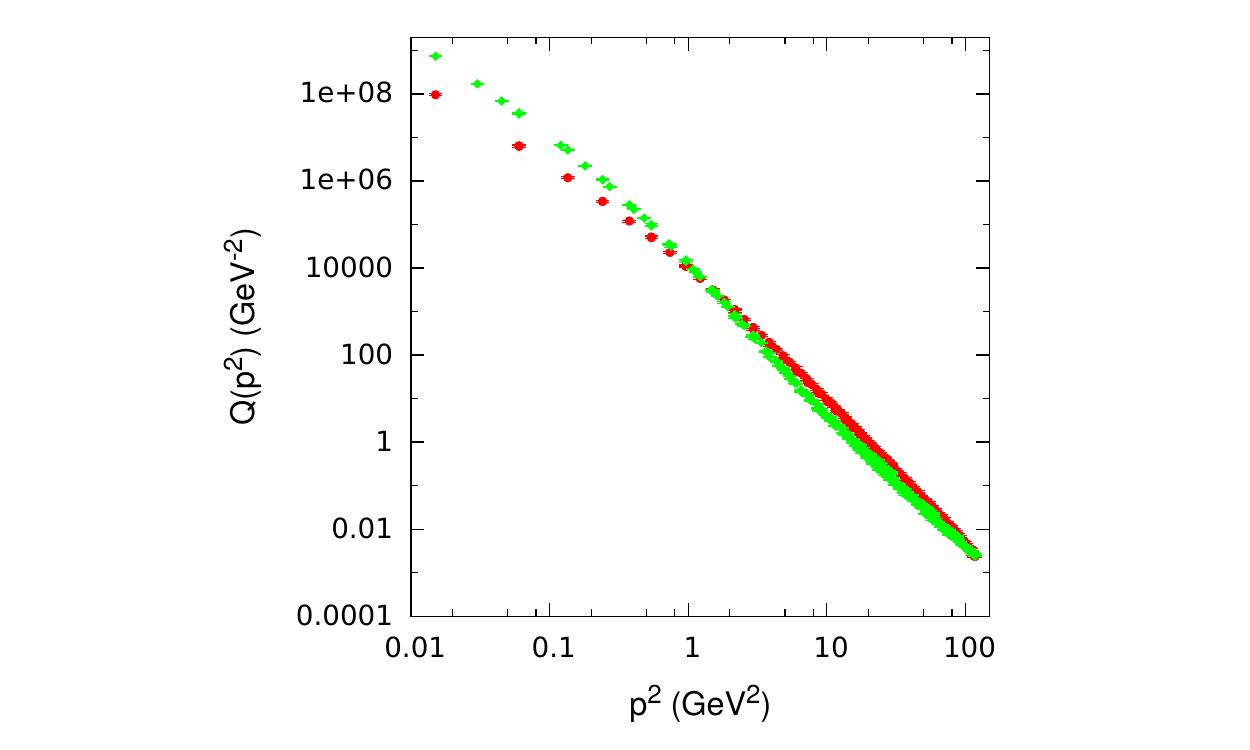}
\caption{
\label{fig:props}
The Bose-ghost propagator
$Q(p^2)$ (\protect\marksymbol{*}{red}) and the product
$g_0^2 \, G^2(p^2) \, D(p^2)$ (\protect\marksymbol{diamond*}{green})
as a function of the (improved) lattice momentum
squared $p^2$
for the lattice volume $V = 120^4$
at $\beta_3 \approx 2.51$.
Here we used as sources $B^{ec}_{\mu}(z)$
the formula reported in Eq.\ (32) of Ref.\ \cite{Cucchieri:2016czg}.
Note the logarithmic scale on both axes.
}
\end{center}
\end{figure}

In Figs.\ \ref{fig:props-beta0} and
\ref{fig:props} we compare the Bose-ghost propagator
$Q(p^2)$ to the product $g_0^2 \, G^2(p^2) \, D(p^2)$, where
$g_0$ is the bare coupling constant. To this end, the
data of the Bose-ghost propagator have been
rescaled in order to agree with the data of the
product $g_0^2 \, G^2(p^2) \, D(p^2)$
at the largest momentum.\footnote{This is equivalent to
imposing a given renormalization condition for the propagators
at the largest momentum.}
This comparison is based on the result
\begin{equation}
Q(p^2) \, \sim \, g_0^2 \, G^2(p^2) \, D(p^2)
   \; ,
\label{eq:QpropDG}
\end{equation}
obtained in Ref.\ \cite{Zwanziger:2009je}
using a cluster decomposition.
Even though there is a clear discrepancy between these two
quantities we find that this discordance seems to
decrease when the continuum limit is considered.

\end{document}